# Doping dependence of the upper critical field, superconducting current density and thermally activated flux flow activation energy in polycrystalline CeFeAsO$_{1-x}$F$_x$ superconductors

S.V. Chong [a,b,*], G.V.M. Williams [b,c], S. Sambale [a,c], K. Kadowaki [d]

[a] *Robinson Research Institute,Victoria University of Wellington, PO Box 33436, Lower Hutt 5046, New Zealand*

[b] *MacDiarmid Institute for Advanced Materials and Nanotechnology, Victoria University of Wellington, PO Box 600, Wellington 6140, New Zealand*

[c] *School of Chemical and Physical Sciences, Victoria University of Wellington, PO Box 600, Wellington 6140, New Zealand*

[d] *Division of Materials Science, Faculty of Pure & Applied Sciences, University of Tsukuba, 1-1-1, Tennodai, Tsukuba, Ibaraki 305-8573, Japan*

**Abstract**

We report the results from resistivity and magnetic measurements on polycrystalline Ce oxypnictide (CeFeAsO$_{1-x}$F$_x$) samples where $x$ spans from 0.13 to 0.25. We find that the orbital limiting field is as high as 150 T and it systematically decreases with increasing doping. The Maki parameter is greater than one across the phase diagram and the large Maki parameter suggests that orbital and Pauli limiting effects contribute to the upper critical field. The broadening of the superconducting transition in the resistivity data was interpreted using the thermally activated flux flow (TAFF) model where we find that the TAFF activation energy, $U_0(B)$, is proportional to $B^{-\gamma}$ from 1 T to high fields, and $\gamma$ does not significantly change with doping. However, $U_0$ and the superconducting critical current, $J_c$, are peaked in the mid-doping region ($x = 0.15$ to $x = 0.20$), and not in the low ($x < 0.15$) or high doping ($x > 0.20$) regions. Furthermore, $U_0$ is correlated with $J_c$ and follows the two fluid model for granular samples.

* Corresponding author. Tel.: +64 4 463 0072.
E-mail address: shen.chong@vuw.ac.nz (S.V. Chong).

# 1. Introduction

The discovery of iron-based compounds with superconducting transition temperatures, $T_c$'s, as high as 56 K, second only to the high temperature superconducting cuprates (HTSCs), has provided us with a new candidate for technological applications using superconducting materials. Iron-based superconductors possess several appealing features which make them a strong candidate for high magnetic field applications at low temperature. With $T_c$'s comparably higher than conventional low-temperature superconductors, such as $Nb_3Sn$, this new class of superconductors also has a very high upper critical field, $B_{c2}$, and high critical currents [1]. $B_{c2}$ of the iron-based superconductors was initially reported to be large in the 1111- and 122-Fe arsenides, and subsequently a large $B_{c2}$ (with $dB_{c2}/dT > $ -5 T/K) was also reported in the 122- and 11-Fe chalcogenides family [2-4]. For example, a $dB_{c2}/dT$ of -12 T/K was recently reported in Rb-doped $TlFe_{2-\delta}Se_2$ with $B \parallel ab$-plane of the single crystal [5], and a more anisotropic $(Sm_{1-x}Nd_x)$-based 1111 superconductor showed a similar value of -11.2 T/K [6]. The high $B_{c2}$ values found in the iron-based superconductors mean that they have potential device applications and hence it would be particularly useful to know how $B_{c2}$ changes with doping. Furthermore, the doping dependence of the superconducting critical current is also required as well as the important role of flux pinning. One study by Shahbazi *et al.* on polycrystalline $CeFeAsO_{1-x}F_x$ with $x = 0.10$ and $x = 0.20$ found that the thermally activated flux flow (TAFF) activation energy, $U_0$, was lower for $x = 0.2$ while the inductive superconducting critical current, $J_c$, was slightly higher for $x = 0.2$ [7]. Thus, there is a clear need to study how $B_{c2}$, $J_c$, and $U_0$ change with doping.

In this paper we report the results from resistivity and magnetic measurements on CeFeAsO$_{1-x}$F$_x$ with nominal $x = 0.13$, $0.15$, $0.20$ and $0.25$. We show that $B_{c2}$ systematically decreases with increasing doping and $U_0$ and $J_c$ are correlated, where $J_c$ has a maximum in the mid-doping region from $x = 0.15$ to $x = 0.20$.

## 2. Experimental details

A series of CeFeAsO$_{1-x}$F$_x$ polycrystalline samples was prepared by a two-step solid state reaction as previously described [8,9]. The first step involved reacting a stoichiometric mixture of CeAs, CeO$_2$, CeF$_3$ and Fe at 1273 K for 24 hours in sealed evacuated quartz ampoules. The reacted mixture was then re-ground, compacted into pellets and again sealed in evacuated quartz ampoules for a final sintering at 1453 K for 50 hours. Powder x-ray diffraction (XRD) of the resulting samples showed the appearance of impurities such as FeAs, FeAs$_2$ and CeAs which increases with fluoride-doping (F-doping) but the total amount is less than 15 % in the highest fluoride-doped sample. The resistivity was measured using a *Quantum Design* Physical Property Measurement System (PPMS) using the four terminal method. The inductive critical current density was determined from magnetization loop (*M-B*) measurements on rectangular-shaped samples with thickness < width << length on a *Quantum Design* Magnetic Property Measurement System (MPMS).

## 3. Results and discussion

The temperature-dependent resistivity at different applied magnetic fields (0 to 8 T) for the F-doped Ce oxypnictide samples is shown in Fig. 1. The increase in $T_c$ is consistent with the doping phase diagram of $CeFeAsO_{1-x}F_x$ reported in literature [10-12], which unlike the cuprates [13] and the electron-doped $BaFe_{2-x}Co_xAs_2$ superconductors [14] do not display a superconducting dome-shape doping phase diagram, and with no downturn in the value of $T_c$ being observed even up to the heavily-doped region ($x = 0.30$) [12].

In the presence of an applied magnetic field the superconducting transition temperature is seen to reduce systematically with increasing magnetic field. Comparing the change in $T_c$ ($\Delta T_c = |T_c^{8T} - T_c^{0T}|$) for all four samples, it is observed in Fig. 1 that $\Delta T_c$ increases with increasing F-doping with $\Delta T_c = 1.03$ K for $x = 0.13$ and it increases to 3.90 K for $x = 0.25$. This indicates that there is a change in $B_{c2}$ with doping. The powder averaged upper critical field was estimated from the temperature where the resistivity had decreased to 90% of the normal state resistivity, which is the method used in the literature [2,3,6,7]. The resultant $B_{c2}$ is plotted in Fig. 2(a) where it is apparent that the gradient is largest for the underdoped sample with $x = 0.13$. This is clearer in Fig. 2(b) where $B_{c2}$ is plotted against the reduced temperature, $T/T_c$. Similar to another study on polycrystalline $CeFeAsO_{1-x}F_x$, we find that $B_{c2}$ is not linear over the entire magnetic field range and it deviates from the high field linearity close to $T_c$ [7]. Thus, we use the average gradient above 1 T to estimate $dB_{c2}/dT$ from the data in Fig. 2, similar to the "linear analysis" used to obtain $B_{c2}$ in the $YBa_2Cu_3O_{7-\delta}$ [15] and $MgB_2$ [16] superconductors. The resultant $dB_{c2}/dT$ is plotted in Fig. 3(a) against $x$ where we find that $dB_{c2}/dT$ is -6.1 T/K for the 0.13 F-doped sample and decreases to -2.1 T/K for the 0.25 F-doped sample. These estimated slopes are as high as those reported by Shahbazi *et al.* [7] for the 10% and the 20% F-doped samples where $dB_{c2}/dT$ has values of -5.9 and -2.4 T/K, respectively, but they are overall higher than those reported by Prakash *et al.* [17] with $dB_{c2}/dT$ values of -3.52 and -1.45 T/K for 10% and the 20% F-doping, respectively.

Similar to other studies [5,17,18], we estimate the orbital limiting field at 0 K, $B_{c2}^{orb}(0)$, using the single-band Bardeen-Cooper-Scrieffer (BCS) Werthamer-Helfand-Hohenberg (WHH) formula [19], $B_{c2}^{orb}(0) = 0.693T_c|dB_{c2}/dT|_{Tc}$. The resultant $B_{c2}^{orb}(0)$ values are plotted in Fig. 3(b) based on the $dB_{c2}/dT$ values in Fig. 3(a). We find that $B_{c2}^{orb}(0)$ decreases with increasing $x$ and $B_{c2}^{orb}(0)$ is nearly independent of doping for $x \geq 0.20$.

It has previously been noted from studies on a number of iron-based superconductors that $B_{c2}(0)$ can exceed the Pauli limiting field, $B_P(0)$, expected in a BCS superconductor [3,20,21]. The Pauli limiting field can be defined as the magnetic field where the superconducting condensate energy is equal to the normal state Zeeman energy [22-24]. This leads to a BCS $B_P(0)$ of $B_P(0) = \mu_0 1.84 * T_c$ [23]. The relative importance of the Pauli limiting field in determining the upper critical field in a single-band model can be described using the Maki parameter, $\alpha = \sqrt{2} * B_{c2}^{orb}(0)/B_P(0)$ [22,24]. We show in Fig. 3(b) that $B_{c2}^{orb}(0)$ exceeds $B_P(0)$ for $x \leq 0.15$ and $B_P(0)$ is slightly higher than $B_{c2}^{orb}(0)$ for $x \geq 0.20$. The resultant $\alpha$ is plotted in the inset to Fig. 3(b) where we find that $\alpha$ is close to 1 for $x \geq 0.20$, and its value becomes significantly greater than 1 for $x \leq 0.15$. The large $\alpha$ values suggest that the pair-breaking includes orbital as well as Pauli limiting effects [22]. Maki parameters greater than 1 have been observed in other $R$FeAsO superconductors with $R$ = La [20] and $R$ = Sm [21] and they may suggest that spin-orbit scattering should be included in the calculation of $B_{c2}$ [24]. However, the situation is more complicated in the iron-based superconductors because there is growing evidence that a two-gap model is required to describe $B_{c2}$ [20,21,25,26] although this can depend on the orientation of the *ab*-plane with respect to the applied magnetic field [21]. Irrespective of the $B_{c2}$ model, the decrease in -d$B_{c2}$/d$T$ with increasing $x$ (Fig 3(a)) suggests that $B_{c2}(0)$ is lower in the high doping side of the phase diagram.

It is apparent in Fig. 1 that the superconducting transition becomes broader as the magnetic field is increased. This is commonly observed in the iron-pnictides [7,27,28] and HTSCs [29-31], and it is interpreted in terms of vortex motion due to the thermal activation of vortices from their respective pinning wells at temperatures where the thermal energy becomes comparable to the pinning potential causing dissipation in the superconductor. This is the thermally activated flux-flow (TAFF) or creep of vortices where the low temperature resistivity is modelled as, $\rho(T,B) = \rho_{0}*\exp(-U_{0}/k_{B}T)$ [3,32-34], where $\rho_{0}$ is the temperature-independent resistivity, $k_{B}$ is Boltzmann's constant, and $T$ is the temperature. The TAFF activation energy, $U_{0}$, was calculated from the resistivity data and for temperatures from the mid-point $T_{c}$ and below and this corresponds to the TAFF region where an Arrhenius plot yields a straight line, and $U_{0}$ is the slope of the linear part of the $\ln(\rho)$ versus $T^{-1}$ plot.

The magnetic field dependence of $U_{0}$ is plotted in Fig. 4. It is apparent that $U_{0}$ is largest around the mid-doping region ($x = 0.15$ to $x = 0.20$) which is clearer in Fig. 5 (filled circles, left axis) where $U_{0}$ is plotted at 1 T. We show in Fig. 4 that $U_{0}(B)$ has a power-law dependence on the magnetic field, $U_{0} \propto B^{-\gamma}$, with $\gamma$ in the range of 0.41 to 0.50 for $B > 1$ T (see the inset to Fig. 4). A $\gamma$ value of 0.5 and above at high field (> 1 Tesla) has also been reported for P-doped PrOFe$_{0.9}$Co$_{0.1}$As ($\gamma \sim 0.5$-0.8) [35], FeTe$_{0.60}$Se$_{0.40}$ ($\gamma \sim$ 0.57 for $B > 6$ T) [28] and CeFeAsO$_{1-x}$F$_{x}$ ($\gamma = 0.7$ for $B > 3$ T) [7]. Large $\gamma$ values have been interpreted in terms of collective creep rather than a single flux line response [18,24,36]. The latter (single flux line) has very weak field-dependent behaviour for $U_0$ and only occurs in the low field region ($< \sim 1$ T). High $\gamma$ values can also occur via plastic deformation of the flux line lattice at dislocations such as those observed in some HTSCs [37]. The range of $U_0(B)$ values at 1 Tesla is similar to those found in Ce-based 1111 ($U_0(1$ T$) \sim 426$-1374 K) [7] and La-based 1111 ($U_0(1$ T$) \sim 662$-800 K) [7], but is much smaller than that found in Sm-based 1111 ($U_0(1$ T$) \sim 5605$-12482 K) [38,39] and Nd-based 1111 ($U_0(1$ T$) \sim 1794$-9804 K) [32] superconductors, indicating a lower pinning force for La- and Ce-based 1111 superconductors.

The critical current density, $J_c$, was determined from the width of irreversible magnetization $\Delta M = (M^- - M^+)$, where $M^+$ and $M^-$ are the branches of the magnetization field-loop ($M$-$B$) for increasing and decreasing applied magnetic field, respectively. The paramagnetic component ($M_p \approx (M^- + M^+)/2$) was first subtracted from the $M$-$B$ loop before using the Bean critical state model [40,41] to obtain $J_c = 3\Delta M/R$ (where $R$ is the average particle size), which is the current flowing only within the grains. The intragrain $J_c$ values obtained for the four F-doped Ce-based 1111 samples at 5 K are plotted in Fig. 6. The average particle size was taken to be $\sim 10$ μm, which was determined from scanning electron microscope imaging.

In all four samples, the field dependence of $J_c$ is similar and can be divided into three different regimes. In the lowest field region ($B < \sim 0.04$ T), $J_c$ is nearly field-independent where the self-field produced by the screening current is much higher than the external magnetic field. Above the self-field region ($B > \sim 0.04$ T), there is an intermediate region where $J_c$ is seen to decrease rapidly, which is commonly observed in both low and high-temperature superconductors [42]. In this intermediate region, $J_c$ has an inverse power-law dependence on $B$, i.e. $J_c(B) \propto B^{-\nu}$, where the log-log plot is linear [43] as indicated by the solid straight lines in Fig. 6. We find that the exponent increases with increasing doping from 0.39 for $x = 0.13$ to 0.62 for $x = 0.25$ as shown in the inset to Fig. 6. This range of exponent values is consistent with that observed in the HTSCs where $v$ ranges from 0.25 to 1 [43,44]. A power-law behaviour is attributed to the existence of networks of weak-links in between grains and it is also believed to be influenced by the thickness of the grain boundaries according to studies done on the HTSCs [42,43].

In the high field region above $B^*$ (see Fig. 6) we find the $J_c$ no longer shows a power-law dependence on the applied magnetic field. This has also been observed in the HTSCs and it indicates that the weak linked regions no longer contribute to $J_c$ [43]. There is also evidence of a peak-effect in $J_c$ that is clear for the $x = 0.15$ sample, which is observed in some HTSCs [15,45] and some iron-based superconductors [7,8,46-49]. There are a number of different models to explain the peak-effect that include the magnetic field dependence of vortex pinning potential and different crossover regimes in the vortex structure [47,49]. Interestingly, we find that $B^*$, which indicates the onset of the peak-effect, is larger for high fluoride doping as can be seen in the inset to Fig. 6.

It is apparent in Fig. 6 that in the entire measured field range, $J_c$ for the 0.15 and 0.20 F-doped samples is higher than the other two F-doped samples. This means that the intragrain $J_c$ for our CeFeAsO$_{1-x}$F$_x$ samples does not change monotonically with doping. This is clearer in Fig. 5 where $J_c$(1 T) is plotted at 5 K and 10 K where $J_c$(1 T) and $U_0$(1 T) are peaked in the mid-doping region ($x = 0.15$ to $x = 0.20$). We also find that the current carrying capacity for $x = 0.15$ and 0.20 is still high and in the $10^5$ A/cm$^2$ range at 6 T (see Fig. 6). Furthermore, the self-field $J_c$(0) at 5 K for the 0.15 and 0.20 F-doped samples is close to $1.5 \times 10^6$ A/cm$^2$, while those of the 0.25 and 0.13 F-doped samples are in the $10^5$ A/cm$^2$ range.

A maxima in $J_c$ with electronic doping was also reported by Tallon *et al.* [50] from measurements on the $Y_{1-x}Ca_xBa_2Cu_3O_{7-\delta}$ HTSC where the peak occurred in the slightly overdoped region. In that case the $J_c$ peak was attributed to a normal state pseudogap in the underdoped region that reduces the superconducting order parameter and increases $J_c$ up until the number of doped holes per Cu, $p$, reaches 0.19 and where the normal state pseudogap has reduced to zero. $J_c$ is believed to fall for $p > 0.19$ due to a reduction in the superconducting gap energy. We are not aware of reports of a normal state pseudogap in $CeFeAsO_{1-x}F_x$ although a pseudogap has been reported in $LaFeAsO_{1-x}F_x$ [51-53] and $SmFeAsO_{1-x}F_x$ [54]. However, it has also been suggested that normal state pseudogap is the cause of the low $B_{c2}$ in $Y_{1-x}Ca_xBa_2Cu_3O_{7-\delta}$ where $B_{c2}$ increases with increasing $p$ up to $p \sim 0.19$ and then decreases [55]. However, as can be seen in Fig. 3, $B_{c2}$ in $CeFeAsO_{1-x}F_x$ decreases with increasing doping, which is opposite to that observed in the HTSCs $Y_{1-x}Ca_xBa_2Cu_3O_{7-\delta}$ [55] and $YBa_2Cu_3O_{7-\delta}$ [15]. Thus, it appears that a pseudogap cannot explain the peak in $J_c$ seen in Fig. 5.

The correlation between $J_c$ and $U_0$ (see Fig. 5) suggests that the peak in $J_c$ arises from a peak in the TAFF $U_0$ and hence it is due to the doping dependence of the depinning energy barrier. A correlation between $U_0$ and $J_c$ is known to occur in granular samples in a two fluid flux creep model where the collective response of vortices operates [18,56]. In this case $U_0(t, B) \approx J_c(0)[Kg(t)t/B]$, where $t = T/T_c$, $g(t) = 4(1-t)^{3/2}$, and $K = (3)^{3/2}\phi_0^2\beta/2c$ ($\phi_0$ = magnetic flux quantum; $c$ = speed of light) [18,56]. By plotting the TAFF $U_0$ at 1 T (Fig. 4) versus the self-field $J_c$ from the 5 K critical current data from Fig. 6, we show in Fig. 7 an excellent correlation between $U_0$ and $J_c$ that is in agreement with the above two fluid flux creep model.

## 4. Conclusions

We have carried out a systematic study of the effect of electron-doping on polycrystalline CeFeAsO$_{1-x}$F$_x$ with nominal $x$ = 0.13, 0.15, 0.20 and 0.25. We find that -d$B_{c2}$/d$T$ systematically decreases with increasing $x$, which suggests that $B_{c2}$(0) also decreases with increasing doping. The orbital limiting field was calculated within the WHH model and it reaches 150 T for $x$ = 0.13. The Maki parameter is slightly greater than 1 for $x \geq 0.20$ and it increases as the doping decreases. The large Maki parameter suggests that Pauli pair-breaking as well as orbital-limiting effects are important across the phase diagram. The TAFF model was used to model the broadening of the resistivity transition. We find that the magnetic field dependence of $U_0(B)$ can be modelled as $U_0 \propto B^{-\gamma}$ from 1 to 8 T, where $\gamma$ does not significantly change with doping. Furthermore, we show that the TAFF activation energy is in the similar range found in other studies on La- and Ce-based 1111 iron pnictides and it is peaked in the mid-doping region ($x$ = 0.15 to $x$ = 0.20). The superconducting critical current density is also peaked around the mid-doping region, and we find that the superconducting critical current is correlated with $U_0$. We find that the superconducting critical current in the intermediate field region follows a power-law dependence where the exponent increases with electron doping. Our results show that device applications using polycrystalline CeFeAsO$_{1-x}$F$_x$ should focus on this mid-doping region because this is where the superconducting critical currents are highest up to the maximum measured magnetic field of 6 T.

## Acknowledgments

This work is supported by the Marsden Fund of New Zealand (VUW0917), the MacDiarmid Institute for Advanced Materials and Nanotechnology, and New Zealand

MBIE (C08X01206). SVC would also like to thank the JSPS postdoctoral fellowship which has funded part of this work.

**Figure captions**:

**Fig. 1.** (color online) Temperature-dependent resistivity of $CeFeAsO_{1-x}F_x$ at different applied magnetic fields. The broadening of the transition with increasing applied field indicates thermally assisted flux flow behaviour.

**Fig. 2.** (color online) (a) Plot of the upper critical fields, $B_{c2}$, against the temperature. (b) Plot of $B_{c2}$ against the reduced temperature, $T/T_c$. The linear lines in (b) are intended to guide the eye.

**Fig. 3.** (color online) (a) Plot of $-dB_{c2}/dT$ against the nominal fluoride concentration, $x$. (b) Plot of the orbital limiting field, $B_{c2}^{orb}(0)$ (filled squares, left axis), and the Pauli limiting field, $B_P(0)$ (filled circles, right axis), against $x$. The inset shows a plot of the Maki parameter, $\alpha$, against $x$.

**Fig. 4.** (color online) Thermally activated flux flow activation energy, $U_0$, versus the applied magnetic field as a function of doping derived from the thermally activated flux flow analysis obtained using the data in Fig. 1. The solid lines are fits to $U_0(B) \propto B^{-\gamma}$. The inset shows the fitted exponent, $\gamma$, against the nominal fluoride concentration, $x$.

**Fig. 5.** (color online) Plot of the thermally activated flux flow activation energy, $U_0$, at 1 T (filled circles, left axis), $J_c$ at 5 K and 1 T (filled squares, right axis), and $J_c$ at 10 K and 1 T (open squares, right axis) against the nominal fluoride concentration, $x$.

**Fig. 6.** (color online) Critical current density $J_c$ versus $B$ at 5 K for CeFeAsO$_{1-x}$F$_x$ with $x$ = 0.13 to 0.25. The solid lines are fits to $J_c(B) \propto B^{-\nu}$. Also shown is $B$* for $x$ = 0.15 and it shows where the power-law region finishes. The inset shows the doping dependent behaviour of the critical field $B$* and the power-law exponent, $\nu$, against the nominal fluoride concentration, $x$.

**Fig. 7.** (color online) Plot of $U_0$(1T) against the self-field $J_c$ at 5 K. The line is a linear fit to the data.

Figure 1

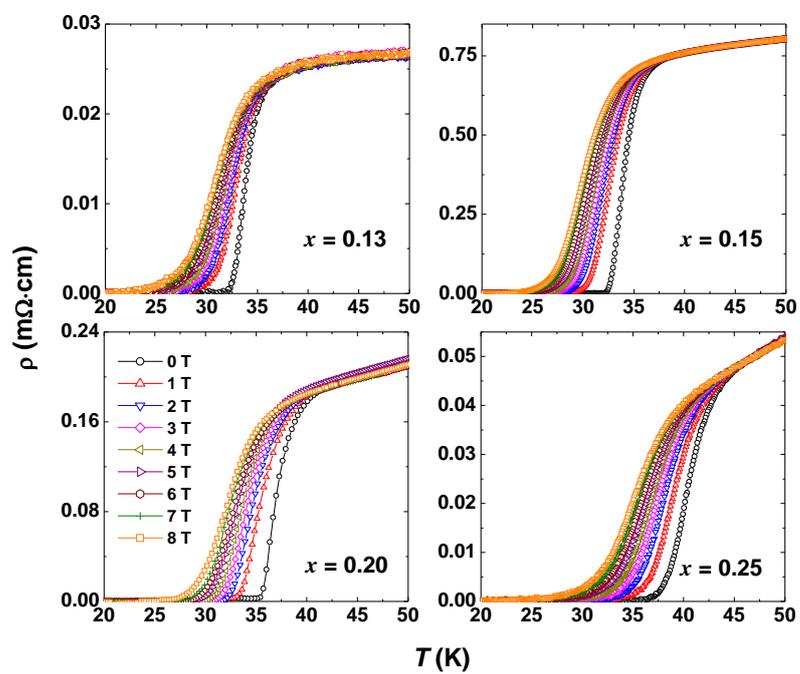

Figure 2

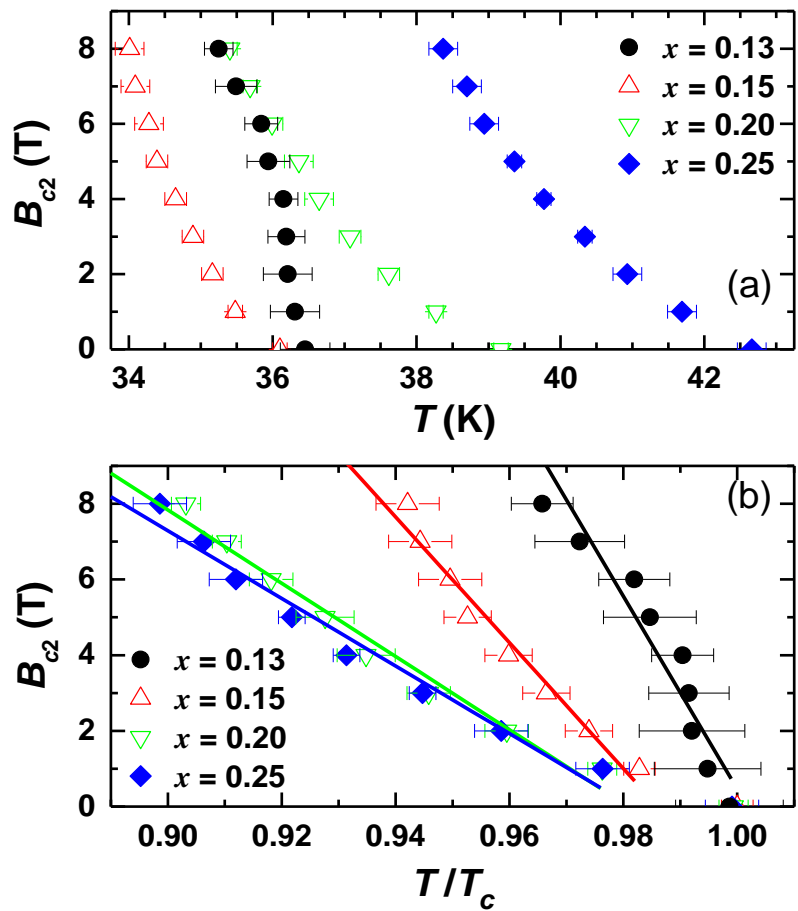

Figure 3

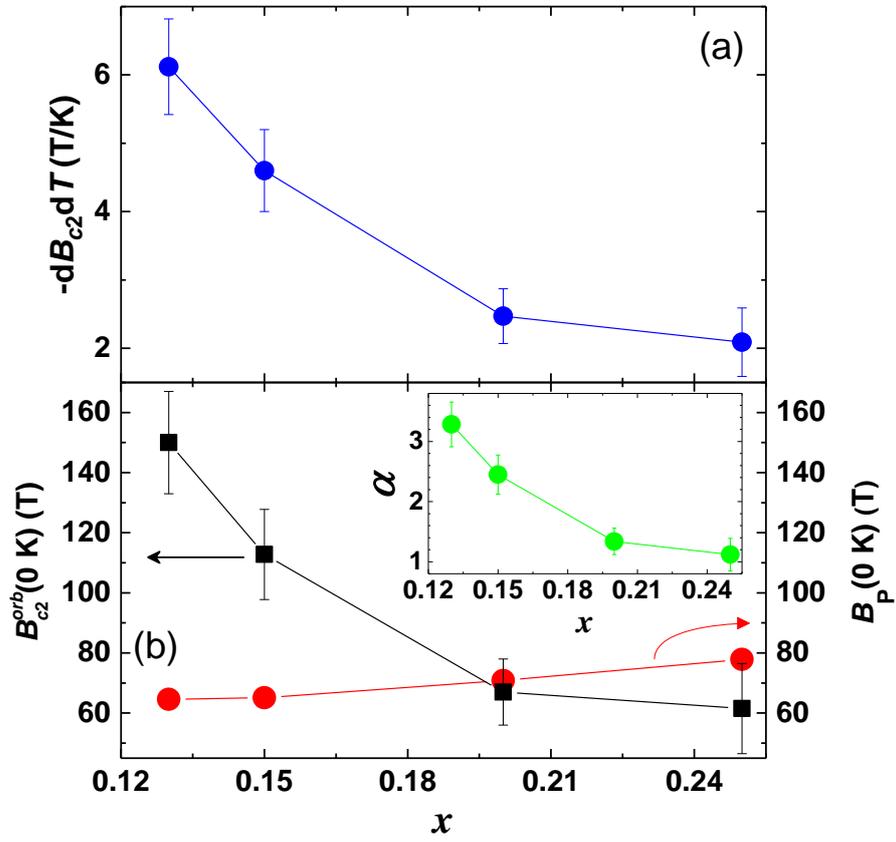

Figure 4

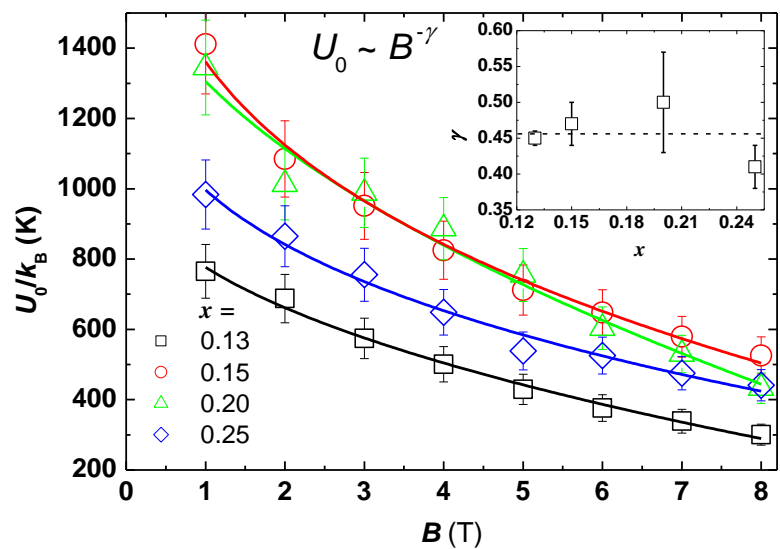

Figure 5

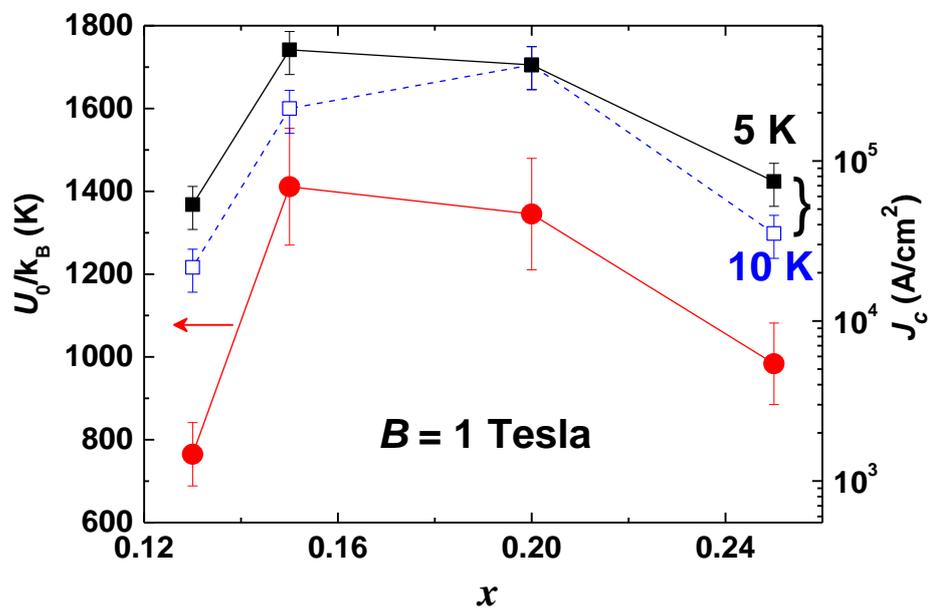

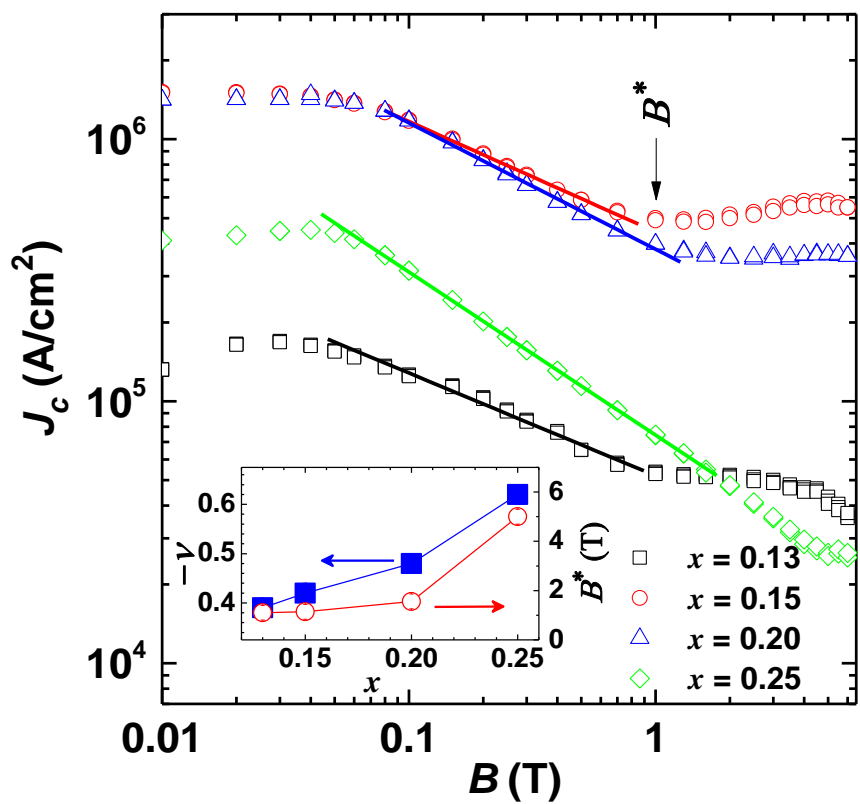

Figure 7

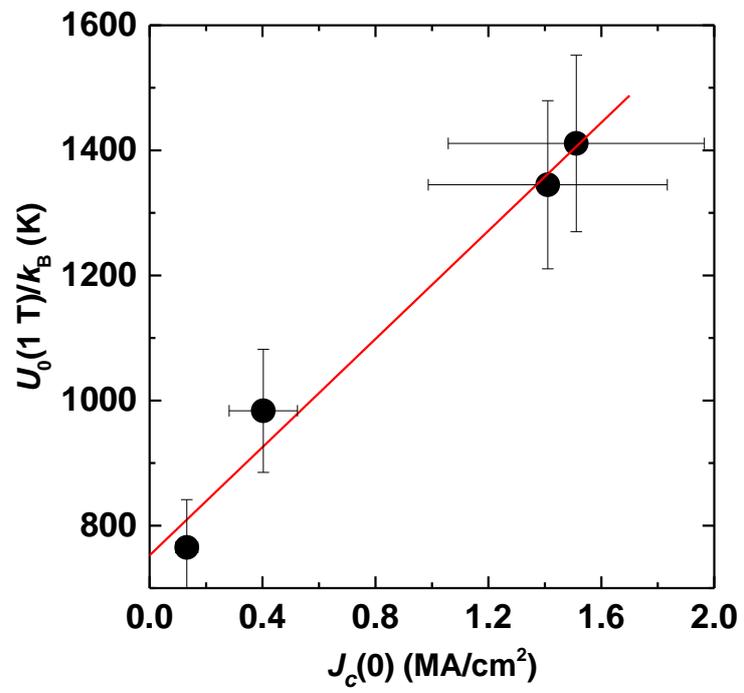